%% file: main.tex
\documentclass{article}

\usepackage[nonatbib,preprint]{neurips_2023}

\usepackage[utf8]{inputenc} 
\usepackage[T1]{fontenc}    
\usepackage{hyperref}       
\usepackage{url}            
\usepackage{booktabs}       
\usepackage{amsfonts}       
\usepackage{nicefrac}       
\usepackage{microtype}      
\usepackage{xcolor}         
\usepackage{microtype}
\usepackage{graphicx}
\usepackage{subfigure}
\usepackage{booktabs} 

\usepackage{hyperref}

\usepackage{amsmath}
\usepackage{amssymb}
\usepackage{mathtools}
\usepackage{amsthm}

\usepackage[capitalize,noabbrev]{cleveref}

\theoremstyle{plain}

\theoremstyle{definition}

\theoremstyle{remark}

\usepackage[textsize=tiny]{todonotes}

\usepackage[utf8]{inputenc}
\usepackage{amsmath,amsfonts,amssymb}
\usepackage[T1]{fontenc}    
\usepackage{hyperref}       
\usepackage{url}            
\usepackage{nicefrac}       
\usepackage{microtype}      
\usepackage{xcolor}         

\usepackage{algorithm}
\usepackage{algpseudocode}
  \DeclareGraphicsExtensions{.png,.pdf}
\usepackage[symbol]{footmisc}
\usepackage{authblk}
\usepackage{physics}
\usepackage{multirow}
\usepackage{longtable}
\usepackage{upquote}

\usepackage{hyperref}
\hypersetup{
    colorlinks=true,
    linkcolor=blue,
    filecolor=magenta,      
    urlcolor=cyan,
    pdftitle={Overleaf Example},
    pdfpagemode=FullScreen,
    }

\usepackage{listings}
\usepackage{color}
\usepackage{dsfont}

\definecolor{dkgreen}{rgb}{0,0.6,0}
\definecolor{dkdkgreen}{rgb}{0,0.3,0}
\definecolor{gray}{rgb}{0.5,0.5,0.5}
\definecolor{mauve}{rgb}{0.58,0,0.82}

\usepackage[numbers]{natbib}
\input{./macros}

\title{Evaluating Physically Motivated Loss Functions for Photometric Redshift Estimation}

\author{%
$\textbf{Andrew Engel}^1 \quad \textbf{Jan Strube}^{1,2}$\\
  $^1$Pacific Northwest National Laboratory\\
  $^2$University of Oregon, Institute of Fundamental Physics\\
  \texttt{\{andrew.engel,jan.strube\}@pnnl.gov}\\
}

\begin{document}

\maketitle

\begin{abstract}
    Physical constraints have been suggested to make neural network models more generalizable, act scientifically plausible, and be more data-efficient over unconstrained baselines. In this report, we present preliminary work on evaluating the effects of adding soft physical constraints to computer vision neural networks trained to estimate the conditional density of redshift on input galaxy images for the Sloan Digital Sky Survey. We introduce physically motivated soft constraint terms that are not implemented with differential or integral operators. We frame this work as a simple ablation study where the effect of including soft physical constraints is compared to an unconstrained baseline. We compare networks using standard point estimate metrics for photometric redshift estimation, as well as metrics to evaluate how faithful our conditional density estimate represents the probability over the ensemble of our test dataset. We find no evidence that the implemented soft physical constraints are more effective regularizers than augmentation.
\end{abstract}

\section{Introduction}
Empirical photometric redshift algorithms regress a target galaxy redshift from measurements of flux received on a telescope's band-pass filters. Highly scalable algorithms for photometric redshifts are critical for achieving the scientific objectives of many upcoming surveys, including the Vera C. Rubin Legacy Survey of Space and Time \cite{LSST_doc}, the Euclid Wide Survey \cite{Eulid}, and the Nancy Roman Space Telescope Wide Area Survey \cite{RomanDoc}. Utilizing legacy survey measurements of spectroscopic redshift as labels, we can cast the problem in a supervised machine learning (ML) framework \cite{ANNzCollister2003,Pasquet,CapsNetDey2021,ContrastiveLearningHayat21,CNN_multichannel_PZ_DIsanto2018}. Recent work that has explored ML for photometric redshifts has focused on how to model and evaluate estimates of the conditional density estimate of redshift \cite{CDElossIzbicki2016,CDEloss_software_unused,RecalibratingDey2021}. See \citet{ReviewNewman2022} for a review.

This brief workshop paper builds on work utilizing deep learning computer vision models to regress photometric redshift conditional density estimates \cite{CNN_multichannel_PZ_DIsanto2018,FirstCNNHoyle2015,CNNSDSSLargeHenghes2021,CapsNetDey2021,ContrastiveLearningHayat21} for galaxies with photometry matching the SDSS main galactic sample \cite{SDSS_MGS_2002} and with redshift $\redshift \leq 0.4$ by incorporating soft physical constraints. These are terms added to a loss function that encode known behavior that solutions should exhibit \cite{CharacteringPINNFailureKrishnapryanMahoney2021}. Our main motivation for including these constraints is to evaluate whether they lead to a more robust model (as suggested in \cite{PINNs,PICVBanerjee2023}), here measured using generalization performance on a held-out portion of our dataset. A secondary motivation is to analyze a broad set of soft physical constraints for neural network models to explore if they can be incorporated for scientific objectives. 

\section{Background}

\paragraph{Neural Networks} 
A set of data inputs and targets $\mathcal{D} = \{(\bx_{1},\hat{\bz}_1),(\bx_{2},\hat{\bz}_2),\ldots,(\bx_N,\hat{\bz}_N)\}$ is sampled from a population $\mathcal{X} \subseteq (\R^{N \times m}$,$\R^{N \times 1})$, with $N \in \mathbb{Z}$ number of samples and input feature dimension $m \in \mathbb{Z}$. A neural network is a vector valued differentiable function that maps input samples to vector targets, $f(\bx_i,\; \boldsymbol{\theta}) \to \redshift_i$, and is parameterized by a vector of learnable weights $\boldsymbol{\theta}$. These are updated to minimize an scalar objective function $\mathcal{L}$, via first order optimizers.
This is known as the standard supervised classification problem. 

\paragraph{Photometric Redshift Conditional Density Estimation}
For this task, we model the probability of redshift given the input features with a neural network $\net \sim P(\redshift|\;\bx_i)$. We frame the problem as supervised classification (this approach was taken by \cite{Pasquet}). The loss, $\mathcal{L}$, is chosen to be the cross entropy loss with a weight decay penalty, 
\begin{equation*}
\mathcal{L}_{\text{CE}} = \frac{1}{N} \sum_{i=1}^N  \sum_{c=1}^C -\log(\frac{\exp(\net^c)}{\sum_{j=1}^C \exp(\net^j)}) \cdot \delta_{c}^{\redshift}+ \gamma_{0} ||\boldsymbol{\theta}||^2.
\end{equation*}
Where $\net^j$ is the $j$-th component of the vector-valued function $\net$, $\delta_{c}^{\redshift}$ is the standard Kronecker delta, and $\gamma_{0}$ is a hyperparameter controlling the relative importance of the weight decay term. We have implicitly binned $\redshift$ into discrete bins represented by the vector output of $\net$, such that $\delta_{c}^{\redshift}$ is 1 if $z_{c} < \redshift \leq z_{c+1}$, and 0 otherwise.


\paragraph{Physically Constrained Neural Networks}
We modify the standard conditional density estimation problem described above by appending additional terms to the cross entropy loss that add additional constraints for the model to obey. In full generality, these additional terms take the form:
\begin{equation*}
\mathcal{L}_{\mathrm{phy}} =  \sum_{j=1}^{M}  \gamma_j \sum_{i=1}^N \mathcal{P}_{j}(f(\bx_i,\;\boldsymbol{\theta}))
\end{equation*}
where $\net$ is a neural network acting on datapoint $\bx_i$, M is the number of soft physical constraint terms, $\mathcal{P}_{j}$ is an additional operator encoding physics, and $\gamma_{j}$ is a scalar hyperparameter controlling the relative scale of the constraint. The total loss function can be described as the addition of the ridge regression term and the constraint(s), $\mathcal{L} = \mathcal{L}_{\text{CE}} + \mathcal{L}_{\mathrm{phy}}$

\section{Physically Informed Constraints}
\label{sec:PhysicalConstaints}
\paragraph{Probabilistic Spectrophotmetric Flux Calculation} \label{sec:Spectrohptometric} The spectrophotometric flux $\Phi$ is the flux observed through a photometric filter $R(\lambda)$ by convolving a spectral energy density $F(\lambda)$ with said filter \cite{ABMagnitudeSystem}. The spectrophotometric flux is given as:
\begin{equation*}
  \Phi(\lambda) = \frac{\int_{\lambda_a}^{\lambda_b} F(\lambda) R(\lambda) \lambda \mathrm{d}\lambda}{\int_{\lambda_a}^{\lambda_b}  c \frac{1}{\lambda} R(\lambda) \mathrm{d}\lambda}
\end{equation*}
let the $\Phi'$ represent the spectrophotometric flux that would be measured from the same spectral flux density but at a redshift given by $\bz_i$. The wavelength of such an observer can be related to the original wavelength by $\lambda' = \frac{1+\bz_i}{1+\redshift_i} \lambda$. We will find it convenient to notate this factor as $\nu := \frac{1+\bz_i}{1+\redshift_i}$. We show in Appendix~\ref{app:Derivation} that: 
\begin{equation*}
\Phi'(\nu) = \frac{\int_{\nu \lambda_a}^{\nu \lambda_b} F(\lambda) R(\lambda) \nu^2 \lambda \mathrm{d}\lambda}{\int_{\nu \lambda_a}^{\nu \lambda_b}  c \frac{1}{\lambda} R(\lambda) \mathrm{d} \lambda}.
\end{equation*}
Our neural network $\net$ is trained to approximate the conditional density of redshift given the input data, $p(\bz_i|\bx_i)$. We can incorporate this CDE by calculating the expectation of $\Phi'$. Interpreting the $c$-th output of $\net$ to represent the probability of redshift $p(\bz_i)$, we have $\mathbb{E}[\Phi'](\net) = \int \Phi(\nu) \net \mathrm{d}\bz.$ From this our first physical constraint can be stated as:
\begin{equation*}
\mathcal{P}_{\text{Flux}}(f(\bx_i,\;\boldsymbol{\theta})) = \frac{|\mathbb{E}[\Phi'](\net) - \Phi'(1)|}{\Phi'(1)}.
\end{equation*}

\paragraph{Invariance to Rotations}
\label{sec:Rotations} ``Invariances'' are constraints that can be interpreted as perturbations over which the model output should be unchanged (the term was coined in the context of generative model discriminators in \citet{InvariancesShah2019}). The cosmological redshift of a galaxy is independent of the orientation of the galaxy in the sky. We define the operator $\Theta(\bx)$ to be the ``rotation'' operator, which takes as input image $\bx$ and randomly rotates and flips $\bx$ about the vertical or horizontal axis, then returns the resulting image. Our network output is constrained to be invariant under these rotations by incorporating the physical constraint 
\begin{equation*}
    \mathcal{P}_{\text{rotation}}(f(\bx_i,\;\boldsymbol{\theta})) = | f(\bx_i,\;\btheta) - f(\Theta(\bx_i),\;\btheta)|^2.
\end{equation*}

\paragraph{Invariance to Background Pixels}
\label{sec:Background}
The cosmological redshift is a property of the galaxy centered in the image and therefore independent of the background pixel values. Let $\mathcal{B}(\bx)$ be an operator that resamples the noise of the background sky in input image $\bx$ and returns the resulting image. We describe the background operator in more detail in appendix \ref{app:ResamplingImplementationNotes}. We can write the constraint of invariance to background as 
\begin{equation*}
\mathcal{P}_{\text{background}}(f(\bx_i,\;\boldsymbol{\theta})) = | f(\bx_i,\;\btheta) - f(\mathcal{B}(\bx_i),\;\btheta)|^2.
\label{eq:background_invariance}
\end{equation*}

\paragraph{Conditional Density Estimate Loss}
\label{sec:CDELoss}
To interpret the output of $\net$ as a conditional density estimate certain properties must met. One such property is that we would expect the set of many observations of the random variable $\mathcal{Z}$ from a given galaxy to follow the distribution $\net$. Because there is only one observed redshift per galaxy, we can not evaluate this property exactly; however, we can evaluate a related value which is equal to the mean squared differences between the estimate of CDE and the true CDE up to a constant of integration \cite{CDElossIzbicki2016}.
\begin{equation*}
    \mathcal{P}_{\text{CDE}}(f(\bx_i,\;\boldsymbol{\theta})) =  \int \net^2 \mathrm{d}z - 2 (\net^{c=\hat{z}})^2
\end{equation*}
This term is referred to as the CDE loss within the photometric redshift literature \cite{CDElossIzbicki2016,ConditionalDensityEstimationToolboxDalmasso2019,RecalibratingDey2021}.

\section{Methodology}
\label{sec:Methods}

\paragraph{Compute}
\label{sec:Compute}
We train all models on a DGX-2 A100 server on a single Nvidia A100 GPU.

\paragraph{Datasets} \label{sec:Datasets} We use the benchmark SDSS galactic redshift dataset first described in \citet{Pasquet} and made available online in \citet{CapsData}\footnote{Data available online at: \url{https://biprateep.de/encapZulate-1/data.html}}. We summarize the preparation here, but would refer the reader to \citet{Pasquet} for details. The dataset is created from SDSS DR12 by selecting all spectroscopically confirmed galaxies that satisfy petroMag\_r < 17.77, which is the same cut-off for target selection in the SDSS main galactic sample \cite{SDSS_MGS_2002}. 64 x 64 pixel cutouts are created in each of the five SDSS photometric bands centered on each galaxy. We place 121,543 (20\%) samples randomly into a held-out ``test'' dataset to measure performance at the end of our study. There are 486,169 samples in our training dataset, from which we randomly select 20,000 galaxies for a ``validation set'' that we can use to monitor performance throughout each training run. 

\paragraph{Models}
\label{sec:Models}
Neural networks are constructed as a series of non-linear learnable functions that are commonly abstracted into ``layers'', with the entire configuration of layers called an ``architecture''. The choice of architecture represents a significant hyperparameter. Previous photometric redshift works evaluated the Inception architecture \cite{Pasquet,CNNSDSSLargeHenghes2021,InceptionSzegedy2015}, the Capsule Network \cite{CapsNetDey2021,CapsuleNetworkSabour2017}, and ResNet50 \cite{ContrastiveLearningHayat21,ResNetHe2015}. For simplicity, we chose to evaluate our model on the ResNet50 architecture provided in the repository of \cite{ContrastiveLearningHayat21}, due to preference for the PyTorch framework \cite{PyTorchPaszke2019}. We provide additional implementation details such as choice of hyperparameters in Appendix~\ref{app:NNdetails}.   

\paragraph{Augmentation}
\label{sec:Augmentation}
The invariance to shifts and background pixels would more classically be used as part of a data augmentation pipeline \cite{Pasquet,ContrastiveLearningHayat21}. To separate the regularizing effect of augmentation from the specific inclusion of the invariance to the loss function, we will train our baseline model with and without rotation and background re-sampling augmentations. In practice, we note that because we already include the rotation and sampling as part of our pipeline, we actually combine the pipeline terms into one term:
\begin{equation*}
\mathcal{P}_{\text{invariances}}(f(\bx_i,\;\boldsymbol{\theta})) = | f(\bx_i,\;\btheta) - f(\Theta(\mathcal{B}(\bx_i)),\;\btheta)|^2.
\label{eq:invariances}
\end{equation*}

\subsection{Metrics}
\label{sec:Metrics}
We will track three point estimate metrics consistent with evaluations from prior work. As a measure of spread, we will report the \textbf{median absolute deviation} $\text{\textbf{MAD}} = 1.4826 \times \mathrm{median}(|\frac{\net - \redshift}{1+\redshift}|)$. We track the \textbf{bias} of the residuals as $\mathrm{bias} = \mathbb{E}[\frac{\net - \redshift}{1+\redshift}]$. Finally, we report the \textbf{catastrophic outlier rate}, $\boldsymbol{\mathcal{O}}$ as the fraction of predictions with scaled residual greater than 0.05. These are the same metrics as reported in \cite{CapsNetDey2021}. Taking the number of datapoints in the test dataset to be $N_{\mathrm{test}}$, and the set of integers up to and including N as $\{1,2,\ldots,N\} = [N]$ then $\mathcal{O} = \frac{100}{N_{\mathrm{test}}} \times \bigl{|} \{i \in [N_{\mathrm{test}}] \vert \; \lVert \frac{\net - \redshift}{1+\redshift}\rVert > 0.05 \} \bigr{|}$.

In addition, we also evaluate the performance of our network in modeling the conditional density estimate. We report the value of the mean conditional density estimate loss evaluated over the held-out test dataset and provide visualizations of the probability integral transform (PIT) \cite{PITAstroPolsterer2016,PITOGDawid1984}. We introduce and present PIT visualizations in appendix~\ref{app:PIT}. 

\section{Results}
\label{sec:Results}
A series of ResNet50 neural networks were trained on a random sample of galaxies cutouts from the SDSS DR12 spectroscopic main galaxy sample catalogue to predict the conditional density estimate of redshift under various types of loss functions. We have three different experimental groups. Our baseline evaluates ResNet50 trained in the manner of \cite{Pasquet} with just the cross-entropy loss function. We train a second baseline in the same manner but include both random resampling of background pixels from the sky-distribution and random flips and rotations as augmentations in our data loading pipeline. Our next group uses all of the physical constraints identified in section~\ref{sec:PhysicalConstaints} in tandem with the cross entropy loss term. We present our results in Table~\ref{tab:MainComparison} alongside contemporary works evaluated on nearly the same dataset.

\begin{table}[!ht]
\centering
\caption{Ablation Performance Metrics And Comparable Works}
\begin{tabular}{lllll|l}

\toprule
\textbf{Model} & \textbf{MAD} & \textbf{Bias} & $\boldsymbol{\mathcal{O}}$ & \textbf{CDE loss} & \textbf{Train Time [h]}\\
\midrule
Baseline w/o Aug. & 1.17e-2 & 7.5e-4 & \textbf{1.48\%} & \textbf{-8.34e-4} & \textbf{7.16}\\ 
Baseline w/ Aug. & 1.17e-2 & 7.1e-4 &  2.22\% & -7.69e-4 & 8.16\\
Physically Constrained & \textbf{1.12e-2} &  \textbf{-1.5e-4} & 1.88\% & -8.04e-4 & 24.52\\
\midrule
\citet{SDSSPZLLRBeck2016} & 1.43e-2 & 1.6e-3 & 2.5\% & * & * \\
\citet{CapsNetDey2021} & 8.98e-3 & 7e-5 & 0.19\% & * & * \\
\citet{ContrastiveLearningHayat21} & 8.25e-3 & 1e-4 & 0.21\% & * & * \\
\citet{Pasquet} & 9.12e-3 & 1e-4 & 0.31\% & * & * \\
\bottomrule
\label{tab:MainComparison}
\end{tabular}
\end{table}

\section{Discussion}
\label{sec:Discussion}

This preliminary work seeks to answer whether a set of physical constraints can be added to a network to increase generalization. Our work is interesting because we study soft physical constraints that are not of the typical form from the physically informed literature-- the operators take neither a differential or integral form. We show that it is possible to add penalty terms sensitive to known properties that photometric redshift estimates should obey as additional terms to the loss function.

Our experiment shows that the augmentations implemented have a positive effect on the point-wise performance of the network, and exhibit a regularizing effect that prevent the network features from collapse \cite{NeuralCollapsePapayan2020} on the training data (see appendix~\ref{app:varianceoutputs}). Our PINN-terms do not meaningfully improve over augmentation, having very little effect overall on the point-estimate performance metrics (table~\ref{tab:MainComparison}), how well calibrated the conditional probability estimates are (appendix~\ref{app:PIT}), how much variance the network exhibits on the augmentations (appendix~\ref{app:varianceoutputs}), or on the overall structure of the photometric redshift vs spectroscopic redshift relationship (appendix~\ref{app:pointperformanceplots}). Worse, they dramatically increase training time. We discuss directions for future work in Appendix \ref{app:FutureWork}.

\pagebreak

\bibliographystyle{abbrvnat}
\bibliography{bibliography}

\pagebreak

\appendix

\section{Acknowledgements}
This work made use of data products from the Sloan Digital Sky Survey data release 12. Funding for SDSS-III has been provided by the Alfred P. Sloan Foundation, the Participating Institutions, the National Science Foundation, and the U.S. Department of Energy Office of Science. The SDSS-III web site is \url{http://www.sdss3.org/}.

SDSS-III is managed by the Astrophysical Research Consortium for the Participating Institutions of the SDSS-III Collaboration including the University of Arizona, the Brazilian Participation Group, Brookhaven National Laboratory, Carnegie Mellon University, University of Florida, the French Participation Group, the German Participation Group, Harvard University, the Instituto de Astrofisica de Canarias, the Michigan State/Notre Dame/JINA Participation Group, Johns Hopkins University, Lawrence Berkeley National Laboratory, Max Planck Institute for Astrophysics, Max Planck Institute for Extraterrestrial Physics, New Mexico State University, New York University, Ohio State University, Pennsylvania State University, University of Portsmouth, Princeton University, the Spanish Participation Group, University of Tokyo, University of Utah, Vanderbilt University, University of Virginia, University of Washington, and Yale University.

We thank Nell Byler and Guatham Narayan for help in preparing this manuscript. A.E. and J.S. were supported by the Open Call Initiative, under the Laboratory Directed Research and Development (LDRD) Program at Pacific Northwest National Laboratory (PNNL).  PNNL is a multi-program national laboratory operated for the U.S. Department of Energy (DOE) by Battelle Memorial Institute under Contract No. DE-AC05-76RLO 1830.


\section{Future Work}
\label{app:FutureWork}
The most obvious extension of this work would be to search over hyper-parameter $\gamma_i$ for each loss-term. We would also like multiple runs with different seeds for each experiment to understand the variance of the performance metrics for each experiment. Finally, prior work on understanding the failures of soft-physical constraints from the perspective of the loss function geometry suggest a methodology to visualize the loss landscape and assess whether it is likely our terms help or hurt, so we could study the loss this way to finalize our conclusion \cite{CharacteringPINNFailureKrishnapryanMahoney2021}, and provide analysis for our terms that may provide better understanding for when we might expect soft-physical constraints to be helpful.


A more reaching next step for work that would extend the narrative line we have constructed would seek to combine Template fitting with ML techniques (see for example previous works like \citet{}). Using existing template libraries \cite{EasyTemplateLibraries}, we could add an additional classification head to the network to choose a template, $F`(\lambda)$ (or even a mixture model of templates). We can then compute the spectrophotometric flux when the template is observed at $\net$ compared to the true spectrophotometric flux observed from the original spectra. The goal would be that our model can be checked against the mixture model of templates classified. 

We also pose the following exploratory analysis to round out the work for a full journal submission: 1) incorporating the unWISE \cite{unWiseLang2014} W1 band and other photometry into the model's MLP head \cite{BeckIRWise2022,CNNSDSSLargeHenghes2021}. 2) Evaluating the model across a sample more representative of the population of galaxies in SDSS's photometric sample rather than the MGS \cite{TeddyBeck2017}. 3) Recent work has made progress pinpointing the effect of mislabeled spectroscopic galaxies on photometric redshift algorithms on simulated datasets \cite{LabelNoiseGPZStylianou2022}. A contamination of just 1\% can cause significant increase in error. A line of research from the interpretability of deep learning research attempt to identify mislabeled samples (sometimes called label-noise in that body of work) through the ``auto-influence'' \cite{InfluenceKoh2017,TracInGarima2020}, which measures how influential a sample is to the loss function (the intuition being at convergence samples which have high residual are most likely mislabeled) This influence can be measured on the scale of our architecture and dataset using projected random matrices techniques \cite{TRAKPark2023}. However, While these methods can help sort which galaxies are mislabeled, they can not actually identify which spectra actually are mislabeled. To quantify the performance, you'd imagine we would need to do the latter on the sample identified, and a large test sample to understand the rate of false positives, and bias of false positives, for such a technique. But we could start by quantifying whether the technique improves the performance metrics with minimal removal of galaxies.   

\section{Full Derivation of Spectrophotometric Flux}
\label{app:Derivation}

If we model the spectrophotometric flux as redshifting the spectra then the two are related as $F(\lambda') = F(\lambda)$. The photometric filter $R(\lambda)$ is defined in the observer's frame and does not receive any redshifting.
With this in place we can write the spectrophotometric flux that an observer at redshift $\net$ from the same galaxy would measure:
\begin{equation*}
  \Phi'(\lambda') = \frac{\int_{\lambda'_a}^{\lambda'_b} F(\lambda') R(\lambda) \lambda' \mathrm{d}\lambda'}{\int_{\lambda'_a}^{\lambda'_b}  c \frac{1}{\lambda'} R(\lambda) \mathrm{d}\lambda'}
\end{equation*}
Making our substitutions and utilizing a change of variables, we can state the spectrophotmetric flux as an equation of the original wavelength and $\nu$. We will drop the argument $\lambda$ as we will show that the crucial dependence of $\Phi'$ on $\net$ is encapsulated in the term $\nu$.
\begin{equation*}
\Phi'(\nu) = \frac{\int_{\nu \lambda_a}^{\nu \lambda_b} F(\lambda) R(\lambda) \nu^2 \lambda \mathrm{d}\lambda}{\int_{\nu \lambda_a}^{\nu \lambda_b}  c \frac{1}{\lambda} R(\lambda) \mathrm{d} \lambda}
\end{equation*}
Because $\net$ approximates a conditional density estimate of $\bz_i$, we can incorporate this CDE by calculating the expectation of $\Phi'$. Interpreting the $c$-th output of $\net$ to represent the probability of redshift $p(\bz_i)$:
$$
\mathbb{E}[\Phi'](\net) = \int \Phi(\nu) \net \mathrm{d}\bz.
$$
From this our first physical constraint can be stated as:
\begin{equation*}
\mathcal{P}_{\text{Flux}} = \frac{|\mathbb{E}[\Phi'](\net) - \Phi'(1)|}{\Phi'(1)}.
\end{equation*}

\section{Limitations} This is preliminary work and all early conclusions that we draw are limited by that fact. Most crucially, we have not performed a hyper-parameter search over values of $\gamma$, nor have we studied the variation in results over multiple training runs initialized form different seeds. For example, this could reveal that in expectation one of the experimental groups could be said to perform better than the others. Our methodology is also limited by the fact that we have not found an efficient way to calculate each of the PINN terms, so even if we found that these PINN terms were to help, its unclear whether small research groups without access to large compute clusters could benefit.






\begin{equation*}
\text{sign}(f-\hat{z}) = \text{sign}((\Phi(f) - \Phi(\hat{z}))\frac{\partial \Phi}{\partial f}) 
\end{equation*}

Typical PINNs do not necessarily align with the $\frac{\partial f}{\partial \theta}$ vector. For example consider the PINN constraint and its derivative:

\begin{equation*}
\mathcal{L}_{\text{PINN}} = |\frac{\partial f}{\partial x}|^2 \\
\nabla_{\theta} \mathcal{L}_{\text{PINN}} = 2 \frac{\partial f}{\partial x} \frac{\partial^2 f}{\partial x \partial \theta} 
\end{equation*}


\section{Various Implementation Notes}
\subsection{Spectrophotometric Loss}
\label{app:FluxImplementationNotes}
Finally, a few notes on the implementation of this computation. We approximate the integrals discretely using the trapezoid rule, which requires that all terms be known at the same $\lambda'$. We linearly interpolate the photometric filter $R(\lambda)$ to achieve this mapping. The entire computation is done inside the PyTorch computational graph, allowing backpropogation through the calculation. One limitation of this technique is that $F(\lambda)$ is only known on the measurement interval of the original measuring spectropraph. If the bandpass filter intersects with a region where the $F(\lambda)$ is outside the measurement range, we set the value of $F(\lambda)$ to zero. This biases our calculation of $\mathbb{E}[\Phi']$. Finally, we avoid the denominator of $\Phi'$ calculation becoming zero by defining $\Phi'$ to be piece-wise equal to 0 when $\int_{\nu \lambda_a}^{\nu \lambda_b}  c \frac{1}{\lambda} R(\lambda) \mathrm{d} \lambda = 0$. A second limitation is that we ignore the measurement error in the spectral flux density, the photometric bandpass filters, and the measured redshift $\redshift$.

\subsection{Rotation Loss}
\label{app:RotationImplementationNotes} As a few notes on our rotational loss implementation, we only evaluate random flips and rotations in increments of $90^{\circ}$, which will preserve the original pixel values to ignore the effects of aliasing. An interesting alternative implementation would be to query for additional pixels than necessary to create our galaxy cutouts which would enable arbitrary rotations without interpolating unknown pixel values. You could ignore the aliasing effects, then randomly draw a rotation angle $\phi$ from the interval $[0,360^\circ)$, and provide this rotation angle to the neural network. Using the rotation angle to rotate the input image as the first layer, it maybe possible to state this constraint as a differential operator $\frac{\mathrm{d}\net}{\mathrm{d}\phi} = 0$, but we leave this to future/ongoing work.

\subsection{Resampling Loss} \label{app:ResamplingImplementationNotes}   As a note on the implementation: separating the background pixels from the galaxy pixel in a manner fast enough to be used as part of the pipeline to a neural network is not a solved problem.  In this iteration of the work we have chosen to trade-off exact modeling of the source for speed in our deep learning pipeline so we do not rely on explicit modeling of the surface brightness of the galaxy. We take a conservative approach instead by setting a threshold value on a background pixel that is equal to the 1$\sigma$ clipped average of the pixels in the image. Pixels that have value less than 1$\sigma$ above the average pixel value are considered background and are randomly drawn from a Gaussian with mean and standard deviation measured from the image. Non-target sources (stars, other nearby galaxies) are not removed by this technique, but it is fast enough to be used in our training pipeline.

\subsection{Neural Network Details}
\label{app:NNdetails}
The neural network utilized is a standard ResNet50 architecture \cite{ResNetHe2015}, re-initialized with random weights from the He Normal initialization. Table~\ref{tab:hyperparameters} gives the hyper-parmaeters for each experimental run. The Adam optimizer \cite{AdamOptimizerKingma2014} was used for each experimental run. For hyperparameters not listed in Table~\ref{tab:hyperparameters} default values as provided in PyTorch 2.0.0 release were used. No dropout was used. The final activation had width 1025, where the extinction due to dust was concatenated into the feature space. 
\begin{table}[!ht]
\centering
\caption{Hyperparameters}
\begin{tabular}{l|lll}
\toprule
\textbf{Hyperparameter} & \textbf{Baseline} & \textbf{Baseline w/ Aug.} & \textbf{PINN} \\
\midrule
$\gamma_{\text{L2}}$ & 1e-06 & 1e-06 & 1e-06  \\
$\gamma_{\Phi}$ & 0 & 0 & 1e-1  \\
$\gamma_{\text{Invariance}}$ & 0 & 0 & 1e-1 \\
$\gamma_{\text{CDELoss}}$ & 0 & 0 & 1e-1 \\
Initial LR & 1e-3 & 1e-3 & 1e-3  \\
Batch Size & 128 & 128 & 128  \\
NEpochs & 50 & 50 & 50  \\
Optimizer & Adam & Adam & Adam \\
LR Schedule factor & 1e-1 & 1e-1 & 1e-1  \\
LR schedule patience & 2 & 2 & 2  \\
\bottomrule
\label{tab:hyperparameters}
\end{tabular}
\end{table}


\section{PIT Visual Metric}
\label{app:PIT}
\begin{figure}
    \centering
    \includegraphics[scale=0.60]{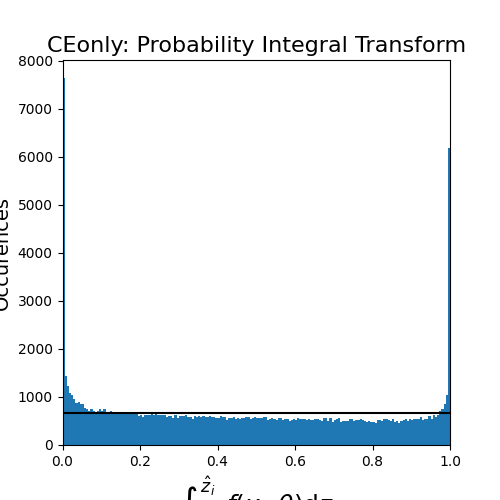}
    \caption{\textbf{Probability Integral Transform of Baseline}} 
    \label{fig:PIT_CE}
\end{figure}

\begin{figure}
    \centering
    \includegraphics[scale=0.60]{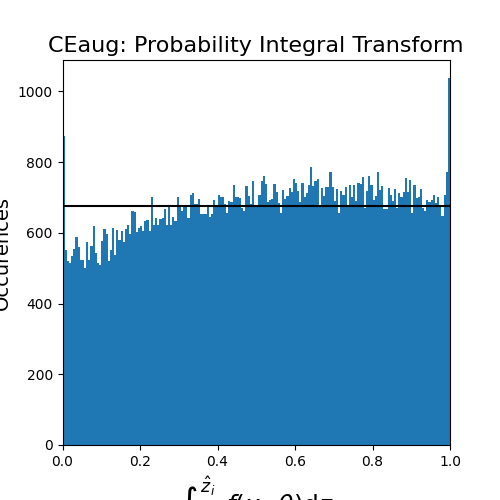}
    \caption{\textbf{Probability Integral Transform of Baseline with augmentations}} 
    \label{fig:PIT_CE_aug}
\end{figure}

\begin{figure}
    \centering
    \includegraphics[scale=0.60]{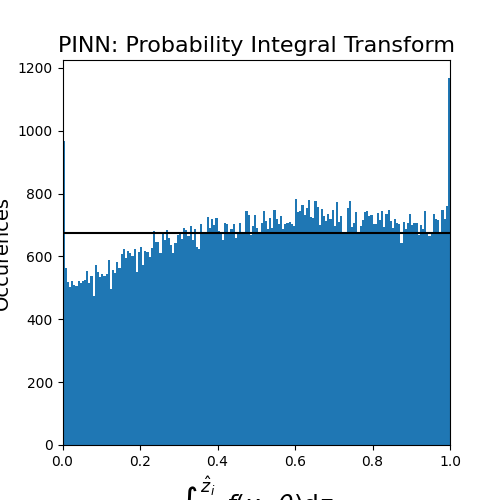}
    \caption{\textbf{Probability Integral Transform of Physically Constrained network}} 
    \label{fig:PIT_PINN}
\end{figure}

The Probability Integral Transform (PIT) is a visual metric that relays information about the bias and how overly or underly-dispersed the probability distributions are. The PIT is a common place metric for photometric redshift evaluation \cite{PZEvaluationLSSTSchmidt2020}. It is simply a histogram of occurrences of cumulative probability distribution integrated up to the true value. The cumulative probability distribution up to the true value is defined:
\begin{equation*}
 CDF_i = \int_{-\infty}^{\redshift} \net \mathrm{d}z.
\end{equation*}
In the below, if the rate of occurrences all fall along the horizontal black line then the probability outputs from the model are well-calibrated. See figures \ref{fig:PIT_CE}, \ref{fig:PIT_CE_aug}, and \ref{fig:PIT_PINN}. 

\section{Visualizing Point-Performance}
\label{app:pointperformanceplots}

\begin{figure}
    \centering
    \includegraphics[scale=0.60]{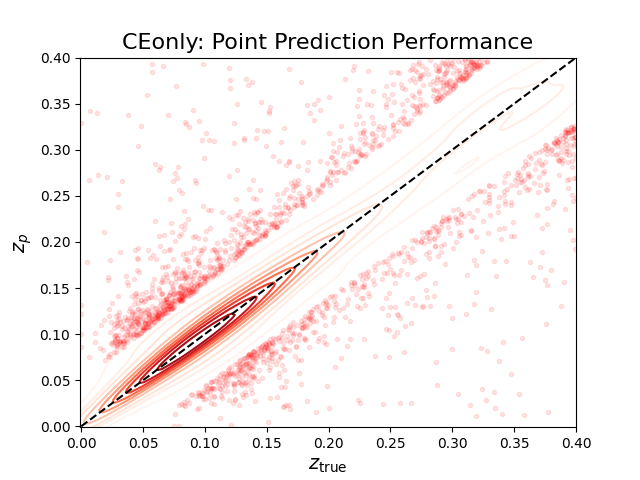}
    \caption{\textbf{KDE of point-estimate performance for Baseline}} 
    \label{fig:ppp_CE}
\end{figure}

\begin{figure}
    \centering
    \includegraphics[scale=0.60]{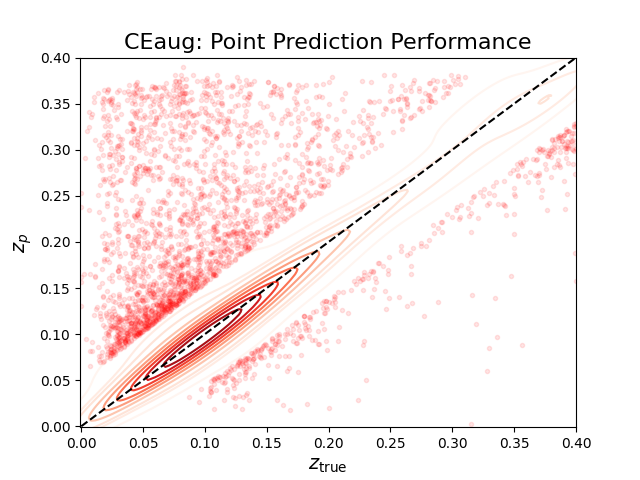}
    \caption{\textbf{KDE of point-estimate performance for Baseline with augmentations}} 
    \label{fig:ppp_CEOnly}
\end{figure}

\begin{figure}
    \centering
    \includegraphics[scale=0.60]{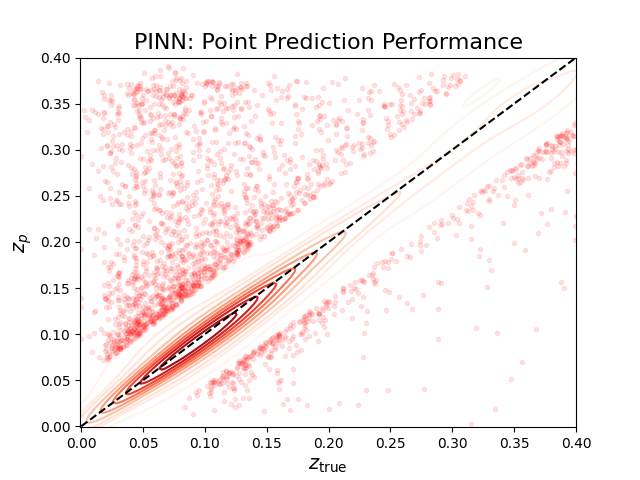}
    \caption{\textbf{KDE of point-estimate performance for PINN experiment}} 
    \label{fig:ppp_PINN}
\end{figure}

In this section we plot the point estimates of photometric redshift from the expectation of the conditional density estimate output from our model against the true spectroscopic redshift using a Kernel Density Estimate (KDE) to visualize the density. We plot all catastrophic outliers as points to visualize their distribution as well. See figures \ref{fig:ppp_CE},  \ref{fig:ppp_CEOnly},  \ref{fig:ppp_PINN}

\subsection{Visualizing Outliers in Point-Performance}

\begin{figure}
    \centering
    \includegraphics[scale=0.60]{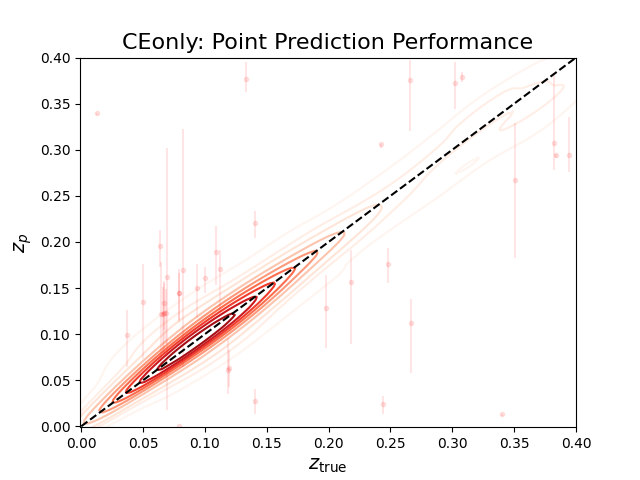}
    \caption{\textbf{KDE of point-estimate performance for Baseline, with random selection of errors}} 
    \label{fig:ppperr_CE}
\end{figure}

\begin{figure}
    \centering
    \includegraphics[scale=0.60]{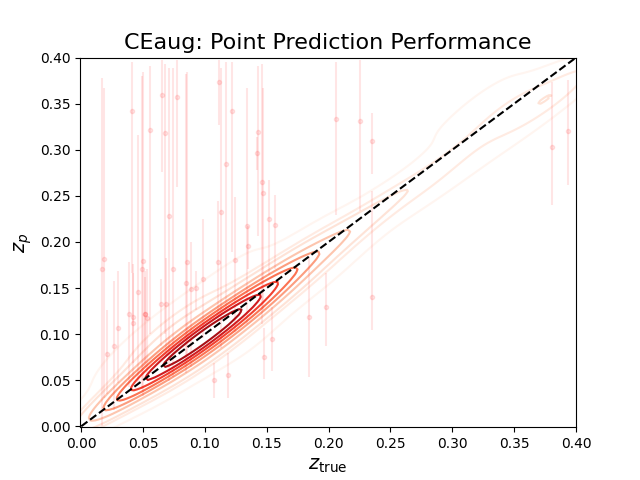}
    \caption{\textbf{KDE of point-estimate performance for Baseline with augmentations, with random selection of errors}} 
    \label{fig:ppperr_CEOnly}
\end{figure}

\begin{figure}
    \centering
    \includegraphics[scale=0.60]{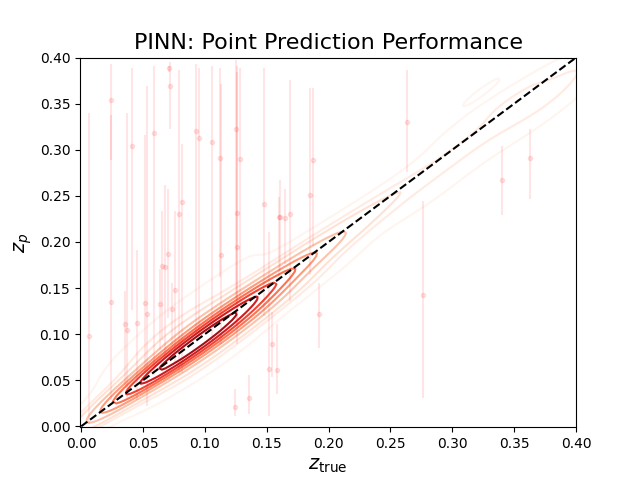}
    \caption{\textbf{KDE of point-estimate performance for PINN experiment, with random selection of errors}} 
    \label{fig:ppperr_PINN}
\end{figure}

In this section we visualize the performance on our test-dataset just as above, but use the error-bar plotting functionality to randomly select a few catastrophic errors and plot their 0.05-0.95 confidence regions are vertical error bars. We compute the 0.05-0.95 confidence regions from the output of our model. We see evidence that many of these catastrophic errors may actually be within the expected tolerance given the output estimate of the conditional density. Future work could investigate the use of these error estimates to flag likely catastrophic errors from the test dataset for downstream cosmological analysis. Previous works have utilized photometric band alone to identify likely catastrophic outliers \cite{IdentifyingOutliersSingal}. See figures \ref{fig:ppperr_CE}, \ref{fig:ppperr_CEOnly}, and \ref{fig:ppperr_PINN}.

\section{Visualizing Predictions and Variability due to Augmentations}
\label{app:varianceoutputs}
\begin{figure}
    \centering
    \includegraphics[scale=0.80]{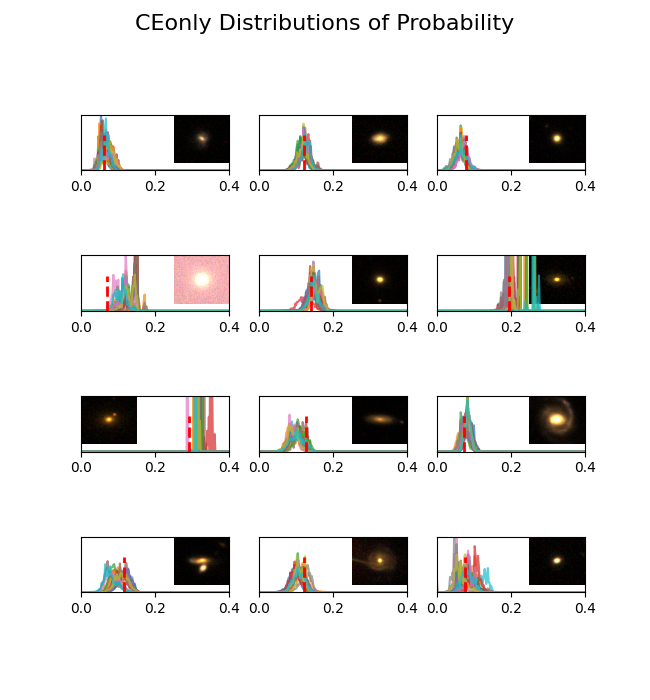}
    \caption{\textbf{A selection of Baseline model output conditional density estimates plot over 20 random rotations and background re-samplings of the shown galaxy.}} 
    \label{fig:dist_CE}
\end{figure}

\begin{figure}
    \centering
    \includegraphics[scale=0.80]{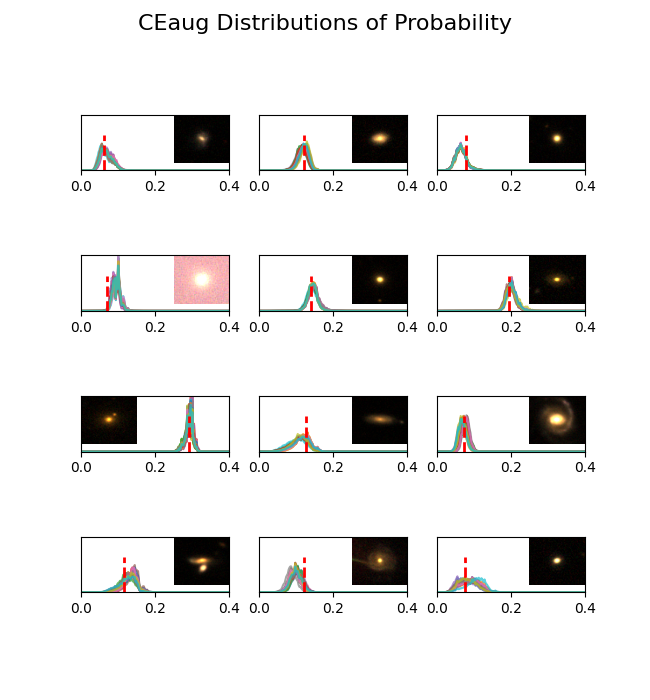}
    \caption{\textbf{A selection of Baseline w/ Augmentation model output conditional density estimates plot over 20 random rotations and background re-samplings of the shown galaxy.}} 
    \label{fig:dist_CEOnly}
\end{figure}

\begin{figure}
    \centering
    \includegraphics[scale=0.80]{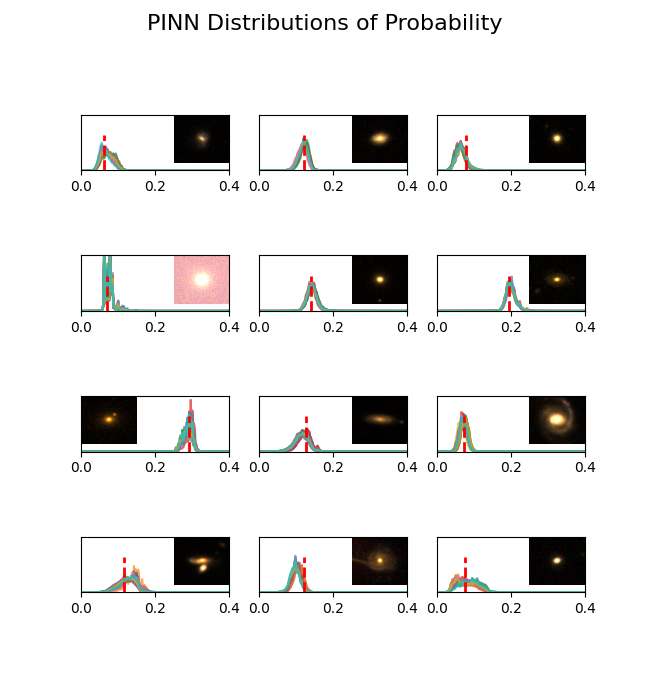}
    \caption{\textbf{A selection of PINN model output conditional density estimates plot over 20 random rotations and background re-samplings of the shown galaxy.}} 
    \label{fig:dist_PINN}
\end{figure}

In this section we randomly select 12 galaxies from the test set and show the distribution of outputs from each model on those same 12 galaxies. We use our augmentation pipeline to randomly add flips and rotations, and resample the background, to produce 20 different estimates of the same galaxy's conditional density of Redshift. The main observation is that the baseline model without augmentation produced more collapsed probability estimates with more variation to these rotations, which we should not expect the network to be sensitive to. See figures \ref{fig:dist_CE}, \ref{fig:dist_CEOnly}, and \ref{fig:dist_PINN}. 

\end{document}

%% file: macros.tex
\def\bx{{\boldsymbol x}}

\def\bz{{\boldsymbol z}}

\def\btheta{{\boldsymbol \theta}}

\newcommand{\R}{\mathbb{R}}

\newcommand{\params}{\boldsymbol{\theta}}
\newcommand{\redshift}{\hat{\bz}}
\newcommand{\inputs}{\boldsymbol{x}}

\newcommand{\net}{f(\inputs_i,\;\params)}

\definecolor{jgGreen}{rgb}{0.0, 0.5, 0.0}

%% file: main.bbl
\begin{thebibliography}{41}
\providecommand{\natexlab}[1]{#1}
\providecommand{\url}[1]{\texttt{#1}}
\expandafter\ifx\csname urlstyle\endcsname\relax
  \providecommand{\doi}[1]{doi: #1}\else
  \providecommand{\doi}{doi: \begingroup \urlstyle{rm}\Url}\fi

\bibitem[{Akeson} et~al.(2019){Akeson}, {Armus}, {Bachelet}, {Bailey},
  {Bartusek}, {Bellini}, {Benford}, {Bennett}, {Bhattacharya}, {Bohlin},
  {Boyer}, {Bozza}, {Bryden}, {Calchi Novati}, {Carpenter}, {Casertano},
  {Choi}, {Content}, {Dayal}, {Dressler}, {Dor{\'e}}, {Fall}, {Fan}, {Fang},
  {Filippenko}, {Finkelstein}, {Foley}, {Furlanetto}, {Kalirai}, {Gaudi},
  {Gilbert}, {Girard}, {Grady}, {Greene}, {Guhathakurta}, {Heinrich},
  {Hemmati}, {Hendel}, {Henderson}, {Henning}, {Hirata}, {Ho}, {Huff},
  {Hutter}, {Jansen}, {Jha}, {Johnson}, {Jones}, {Kasdin}, {Kelly}, {Kirshner},
  {Koekemoer}, {Kruk}, {Lewis}, {Macintosh}, {Madau}, {Malhotra}, {Mandel},
  {Massara}, {Masters}, {McEnery}, {McQuinn}, {Melchior}, {Melton},
  {Mennesson}, {Peeples}, {Penny}, {Perlmutter}, {Pisani}, {Plazas}, {Poleski},
  {Postman}, {Ranc}, {Rauscher}, {Rest}, {Roberge}, {Robertson}, {Rodney},
  {Rhoads}, {Rhodes}, {Ryan}, {Sahu}, {Sand}, {Scolnic}, {Seth}, {Shvartzvald},
  {Siellez}, {Smith}, {Spergel}, {Stassun}, {Street}, {Strolger}, {Szalay},
  {Trauger}, {Troxel}, {Turnbull}, {van der Marel}, {von der Linden}, {Wang},
  {Weinberg}, {Williams}, {Windhorst}, {Wollack}, {Wu}, {Yee}, and
  {Zimmerman}]{RomanDoc}
R.~{Akeson}, L.~{Armus}, E.~{Bachelet}, V.~{Bailey}, L.~{Bartusek},
  A.~{Bellini}, D.~{Benford}, D.~{Bennett}, A.~{Bhattacharya}, R.~{Bohlin},
  M.~{Boyer}, V.~{Bozza}, G.~{Bryden}, S.~{Calchi Novati}, K.~{Carpenter},
  S.~{Casertano}, A.~{Choi}, D.~{Content}, P.~{Dayal}, A.~{Dressler},
  O.~{Dor{\'e}}, S.~M. {Fall}, X.~{Fan}, X.~{Fang}, A.~{Filippenko},
  S.~{Finkelstein}, R.~{Foley}, S.~{Furlanetto}, J.~{Kalirai}, B.~S. {Gaudi},
  K.~{Gilbert}, J.~{Girard}, K.~{Grady}, J.~{Greene}, P.~{Guhathakurta},
  C.~{Heinrich}, S.~{Hemmati}, D.~{Hendel}, C.~{Henderson}, T.~{Henning},
  C.~{Hirata}, S.~{Ho}, E.~{Huff}, A.~{Hutter}, R.~{Jansen}, S.~{Jha},
  S.~{Johnson}, D.~{Jones}, J.~{Kasdin}, P.~{Kelly}, R.~{Kirshner},
  A.~{Koekemoer}, J.~{Kruk}, N.~{Lewis}, B.~{Macintosh}, P.~{Madau},
  S.~{Malhotra}, K.~{Mandel}, E.~{Massara}, D.~{Masters}, J.~{McEnery},
  K.~{McQuinn}, P.~{Melchior}, M.~{Melton}, B.~{Mennesson}, M.~{Peeples},
  M.~{Penny}, S.~{Perlmutter}, A.~{Pisani}, A.~{Plazas}, R.~{Poleski},
  M.~{Postman}, C.~{Ranc}, B.~{Rauscher}, A.~{Rest}, A.~{Roberge},
  B.~{Robertson}, S.~{Rodney}, J.~{Rhoads}, J.~{Rhodes}, J.~{Ryan}, Russell,
  K.~{Sahu}, D.~{Sand}, D.~{Scolnic}, A.~{Seth}, Y.~{Shvartzvald},
  K.~{Siellez}, A.~{Smith}, D.~{Spergel}, K.~{Stassun}, R.~{Street}, L.-G.
  {Strolger}, A.~{Szalay}, J.~{Trauger}, M.~A. {Troxel}, M.~{Turnbull}, R.~{van
  der Marel}, A.~{von der Linden}, Y.~{Wang}, D.~{Weinberg}, B.~{Williams},
  R.~{Windhorst}, E.~{Wollack}, H.-Y. {Wu}, J.~{Yee}, and N.~{Zimmerman}.
\newblock {The Wide Field Infrared Survey Telescope: 100 Hubbles for the
  2020s}.
\newblock \emph{arXiv e-prints}, art. arXiv:1902.05569, Feb. 2019.
\newblock \doi{10.48550/arXiv.1902.05569}.

\bibitem[Banerjee et~al.(2023)Banerjee, Nguyen, Fookes, and
  Karniadakis]{PICVBanerjee2023}
C.~K. Banerjee, K.~Nguyen, C.~Fookes, and G.~E. Karniadakis.
\newblock Physics-informed computer vision: A review and perspectives.
\newblock \emph{ArXiv}, abs/2305.18035, 2023.
\newblock URL \url{https://api.semanticscholar.org/CorpusID:258959254}.

\bibitem[Beck et~al.(2016)Beck, Dobos, Budav'ari, Szalay, and
  Csabai]{SDSSPZLLRBeck2016}
R.~Beck, L.~Dobos, T.~Budav'ari, A.~S. Szalay, and I.~Csabai.
\newblock Photometric redshifts for the sdss data release 12.
\newblock \emph{Monthly Notices of the Royal Astronomical Society},
  460:\penalty0 1371--1381, 2016.
\newblock URL \url{https://api.semanticscholar.org/CorpusID:118709999}.

\bibitem[Beck et~al.(2017)Beck, Lin, Ishida, Gieseke, de~Souza, Costa-Duarte,
  Hattab, and Krone-Martins]{TeddyBeck2017}
R.~Beck, C.-A. Lin, E.~E.~O. Ishida, F.~Gieseke, R.~S. de~Souza, M.~V.
  Costa-Duarte, M.~W. Hattab, and A.~Krone-Martins.
\newblock On the realistic validation of photometric redshifts.
\newblock \emph{Monthly Notices of the Royal Astronomical Society},
  468:\penalty0 4323--4339, 2017.
\newblock URL \url{https://api.semanticscholar.org/CorpusID:118980539}.

\bibitem[{Beck} et~al.(2022){Beck}, {Dodds}, and {Szapudi}]{BeckIRWise2022}
R.~{Beck}, S.~C. {Dodds}, and I.~{Szapudi}.
\newblock {WISE-PS1-STRM: neural network source classification and photometric
  redshifts for WISE{\texttimes}PS1}.
\newblock \emph{Monthly Notices of the Royal Astronomical Society},
  515\penalty0 (4):\penalty0 4711--4721, Oct. 2022.
\newblock \doi{10.1093/mnras/stac1714}.

\bibitem[{Brammer} et~al.(2008){Brammer}, {van Dokkum}, and
  {Coppi}]{EasyTemplateLibraries}
G.~B. {Brammer}, P.~G. {van Dokkum}, and P.~{Coppi}.
\newblock {EAZY: A Fast, Public Photometric Redshift Code}.
\newblock \emph{The Astrophysical Journal}, 686\penalty0 (2):\penalty0
  1503--1513, Oct. 2008.
\newblock \doi{10.1086/591786}.

\bibitem[Collister and Lahav(2003)]{ANNzCollister2003}
A.~Collister and O.~Lahav.
\newblock Annz: Estimating photometric redshifts using artificial neural
  networks.
\newblock \emph{Publications of the Astronomical Society of the Pacific},
  116:\penalty0 345 -- 351, 2003.
\newblock URL \url{https://api.semanticscholar.org/CorpusID:119089041}.

\bibitem[Dalmasso et~al.(2019{\natexlab{a}})Dalmasso, Pospisil, Lee, Izbicki,
  Freeman, and Malz]{CDEloss_software_unused}
N.~Dalmasso, T.~Pospisil, A.~B. Lee, R.~Izbicki, P.~E. Freeman, and A.~I. Malz.
\newblock Conditional density estimation tools in python and r with
  applications to photometric redshifts and likelihood-free cosmological
  inference.
\newblock \emph{Astron. Comput.}, 30:\penalty0 100362, 2019{\natexlab{a}}.
\newblock URL \url{https://api.semanticscholar.org/CorpusID:201698434}.

\bibitem[Dalmasso et~al.(2019{\natexlab{b}})Dalmasso, Pospisil, Lee, Izbicki,
  Freeman, and Malz]{ConditionalDensityEstimationToolboxDalmasso2019}
N.~Dalmasso, T.~Pospisil, A.~B. Lee, R.~Izbicki, P.~E. Freeman, and A.~I. Malz.
\newblock Conditional density estimation tools in python and r with
  applications to photometric redshifts and likelihood-free cosmological
  inference.
\newblock \emph{Astron. Comput.}, 30:\penalty0 100362, 2019{\natexlab{b}}.
\newblock URL \url{https://api.semanticscholar.org/CorpusID:201698434}.

\bibitem[Dawid(1984)]{PITOGDawid1984}
A.~P. Dawid.
\newblock The prequential approach.
\newblock \emph{J. R. Stat. Soc. A}, 2:\penalty0 278--292, 1984.
\newblock URL
  \url{https://people.csail.mit.edu/jrennie/trg/papers/dawid-prequential-84.pdf}.

\bibitem[Dey et~al.(2021{\natexlab{a}})Dey, Andrews, Newman, Mao, Rau, and
  Zhou]{CapsNetDey2021}
B.~Dey, B.~H. Andrews, J.~A. Newman, Y.-Y. Mao, M.~M. Rau, and R.~Zhou.
\newblock Photometric redshifts from sdss images with an interpretable deep
  capsule network.
\newblock \emph{Monthly Notices of the Royal Astronomical Society},
  2021{\natexlab{a}}.
\newblock URL \url{https://api.semanticscholar.org/CorpusID:244954597}.

\bibitem[Dey et~al.(2021{\natexlab{b}})Dey, Newman, Andrews, Izbicki, Lee,
  Zhao, Rau, and Malz]{RecalibratingDey2021}
B.~Dey, J.~A. Newman, B.~H. Andrews, R.~Izbicki, A.~B. Lee, D.~Zhao, M.~M. Rau,
  and A.~I. Malz.
\newblock Re-calibrating photometric redshift probability distributions using
  feature-space regression.
\newblock \emph{ArXiv}, abs/2110.15209, 2021{\natexlab{b}}.
\newblock URL \url{https://api.semanticscholar.org/CorpusID:240070954}.

\bibitem[{Dey} et~al.(2022){Dey}, {Andrews}, {Newman}, {Mao}, {Rau}, and
  {Zhou}]{CapsData}
B.~{Dey}, B.~H. {Andrews}, J.~A. {Newman}, Y.-Y. {Mao}, M.~M. {Rau}, and
  R.~{Zhou}.
\newblock {Photometric redshifts from SDSS images with an interpretable deep
  capsule network}.
\newblock \emph{Monthly Notices of the Royal Astronomical Society},
  515\penalty0 (4):\penalty0 5285--5305, Oct. 2022.
\newblock \doi{10.1093/mnras/stac2105}.

\bibitem[{D'Isanto} and {Polsterer}(2018)]{CNN_multichannel_PZ_DIsanto2018}
A.~{D'Isanto} and K.~L. {Polsterer}.
\newblock {Photometric redshift estimation via deep learning. Generalized and
  pre-classification-less, image based, fully probabilistic redshifts}.
\newblock \emph{Astronomy and Astrophysics}, 609:\penalty0 A111, Jan. 2018.
\newblock \doi{10.1051/0004-6361/201731326}.

\bibitem[Hayat et~al.(2021)Hayat, Stein, Harrington, Lukić, and
  Mustafa]{ContrastiveLearningHayat21}
M.~A. Hayat, G.~Stein, P.~Harrington, Z.~Lukić, and M.~Mustafa.
\newblock Self-supervised representation learning for astronomical images.
\newblock \emph{The Astrophysical Journal Letters}, 911\penalty0 (2):\penalty0
  L33, apr 2021.
\newblock \doi{10.3847/2041-8213/abf2c7}.
\newblock URL \url{https://dx.doi.org/10.3847/2041-8213/abf2c7}.

\bibitem[He et~al.(2015)He, Zhang, Ren, and Sun]{ResNetHe2015}
K.~He, X.~Zhang, S.~Ren, and J.~Sun.
\newblock Deep residual learning for image recognition.
\newblock \emph{2016 IEEE Conference on Computer Vision and Pattern Recognition
  (CVPR)}, pages 770--778, 2015.
\newblock URL \url{https://api.semanticscholar.org/CorpusID:206594692}.

\bibitem[Henghes et~al.(2021)Henghes, Pettitt, Thiyagalingam, Hey, and
  Lahav]{CNNSDSSLargeHenghes2021}
B.~Henghes, C.~Pettitt, J.~Thiyagalingam, T.~Hey, and O.~Lahav.
\newblock Investigating deep learning methods for obtaining photometric
  redshift estimations from images.
\newblock 2021.
\newblock URL \url{https://api.semanticscholar.org/CorpusID:237420789}.

\bibitem[Hoyle(2015)]{FirstCNNHoyle2015}
B.~Hoyle.
\newblock Measuring photometric redshifts using galaxy images and deep neural
  networks.
\newblock \emph{Astron. Comput.}, 16:\penalty0 34--40, 2015.
\newblock URL \url{https://api.semanticscholar.org/CorpusID:62105017}.

\bibitem[{Ivezi{\'c}} et~al.(2019){Ivezi{\'c}}, {Kahn}, {Tyson}, {Abel},
  {Acosta}, {Allsman}, {Alonso}, {AlSayyad}, {Anderson}, {Andrew}, {Angel},
  {Angeli}, {Ansari}, {Antilogus}, {Araujo}, {Armstrong}, {Arndt}, {Astier},
  {Aubourg}, {Auza}, {Axelrod}, {Bard}, {Barr}, {Barrau}, {Bartlett}, {Bauer},
  {Bauman}, {Baumont}, {Bechtol}, {Bechtol}, {Becker}, {Becla}, {Beldica},
  {Bellavia}, {Bianco}, {Biswas}, {Blanc}, {Blazek}, {Blandford}, {Bloom},
  {Bogart}, {Bond}, {Booth}, {Borgland}, {Borne}, {Bosch}, {Boutigny},
  {Brackett}, {Bradshaw}, {Brandt}, {Brown}, {Bullock}, {Burchat}, {Burke},
  {Cagnoli}, {Calabrese}, {Callahan}, {Callen}, {Carlin}, {Carlson},
  {Chandrasekharan}, {Charles-Emerson}, {Chesley}, {Cheu}, {Chiang}, {Chiang},
  {Chirino}, {Chow}, {Ciardi}, {Claver}, {Cohen-Tanugi}, {Cockrum}, {Coles},
  {Connolly}, {Cook}, {Cooray}, {Covey}, {Cribbs}, {Cui}, {Cutri}, {Daly},
  {Daniel}, {Daruich}, {Daubard}, {Daues}, {Dawson}, {Delgado}, {Dellapenna},
  {de Peyster}, {de Val-Borro}, {Digel}, {Doherty}, {Dubois},
  {Dubois-Felsmann}, {Durech}, {Economou}, {Eifler}, {Eracleous}, {Emmons},
  {Fausti Neto}, {Ferguson}, {Figueroa}, {Fisher-Levine}, {Focke}, {Foss},
  {Frank}, {Freemon}, {Gangler}, {Gawiser}, {Geary}, {Gee}, {Geha}, {Gessner},
  {Gibson}, {Gilmore}, {Glanzman}, {Glick}, {Goldina}, {Goldstein}, {Goodenow},
  {Graham}, {Gressler}, {Gris}, {Guy}, {Guyonnet}, {Haller}, {Harris},
  {Hascall}, {Haupt}, {Hernandez}, {Herrmann}, {Hileman}, {Hoblitt}, {Hodgson},
  {Hogan}, {Howard}, {Huang}, {Huffer}, {Ingraham}, {Innes}, {Jacoby}, {Jain},
  {Jammes}, {Jee}, {Jenness}, {Jernigan}, {Jevremovi{\'c}}, {Johns}, {Johnson},
  {Johnson}, {Jones}, {Juramy-Gilles}, {Juri{\'c}}, {Kalirai}, {Kallivayalil},
  {Kalmbach}, {Kantor}, {Karst}, {Kasliwal}, {Kelly}, {Kessler}, {Kinnison},
  {Kirkby}, {Knox}, {Kotov}, {Krabbendam}, {Krughoff}, {Kub{\'a}nek},
  {Kuczewski}, {Kulkarni}, {Ku}, {Kurita}, {Lage}, {Lambert}, {Lange},
  {Langton}, {Le Guillou}, {Levine}, {Liang}, {Lim}, {Lintott}, {Long},
  {Lopez}, {Lotz}, {Lupton}, {Lust}, {MacArthur}, {Mahabal}, {Mandelbaum},
  {Markiewicz}, {Marsh}, {Marshall}, {Marshall}, {May}, {McKercher}, {McQueen},
  {Meyers}, {Migliore}, {Miller}, {Mills}, {Miraval}, {Moeyens}, {Moolekamp},
  {Monet}, {Moniez}, {Monkewitz}, {Montgomery}, {Morrison}, {Mueller},
  {Muller}, {Mu{\~n}oz Arancibia}, {Neill}, {Newbry}, {Nief}, {Nomerotski},
  {Nordby}, {O'Connor}, {Oliver}, {Olivier}, {Olsen}, {O'Mullane}, {Ortiz},
  {Osier}, {Owen}, {Pain}, {Palecek}, {Parejko}, {Parsons}, {Pease},
  {Peterson}, {Peterson}, {Petravick}, {Libby Petrick}, {Petry},
  {Pierfederici}, {Pietrowicz}, {Pike}, {Pinto}, {Plante}, {Plate}, {Plutchak},
  {Price}, {Prouza}, {Radeka}, {Rajagopal}, {Rasmussen}, {Regnault}, {Reil},
  {Reiss}, {Reuter}, {Ridgway}, {Riot}, {Ritz}, {Robinson}, {Roby}, {Roodman},
  {Rosing}, {Roucelle}, {Rumore}, {Russo}, {Saha}, {Sassolas}, {Schalk},
  {Schellart}, {Schindler}, {Schmidt}, {Schneider}, {Schneider}, {Schoening},
  {Schumacher}, {Schwamb}, {Sebag}, {Selvy}, {Sembroski}, {Seppala}, {Serio},
  {Serrano}, {Shaw}, {Shipsey}, {Sick}, {Silvestri}, {Slater}, {Smith},
  {Smith}, {Sobhani}, {Soldahl}, {Storrie-Lombardi}, {Stover}, {Strauss},
  {Street}, {Stubbs}, {Sullivan}, {Sweeney}, {Swinbank}, {Szalay}, {Takacs},
  {Tether}, {Thaler}, {Thayer}, {Thomas}, {Thornton}, {Thukral}, {Tice},
  {Trilling}, {Turri}, {Van Berg}, {Vanden Berk}, {Vetter}, {Virieux},
  {Vucina}, {Wahl}, {Walkowicz}, {Walsh}, {Walter}, {Wang}, {Wang}, {Warner},
  {Wiecha}, {Willman}, {Winters}, {Wittman}, {Wolff}, {Wood-Vasey}, {Wu},
  {Xin}, {Yoachim}, and {Zhan}]{LSST_doc}
{\v{Z}}.~{Ivezi{\'c}}, S.~M. {Kahn}, J.~A. {Tyson}, B.~{Abel}, E.~{Acosta},
  R.~{Allsman}, D.~{Alonso}, Y.~{AlSayyad}, S.~F. {Anderson}, J.~{Andrew},
  J.~R.~P. {Angel}, G.~Z. {Angeli}, R.~{Ansari}, P.~{Antilogus}, C.~{Araujo},
  R.~{Armstrong}, K.~T. {Arndt}, P.~{Astier}, {\'E}.~{Aubourg}, N.~{Auza},
  T.~S. {Axelrod}, D.~J. {Bard}, J.~D. {Barr}, A.~{Barrau}, J.~G. {Bartlett},
  A.~E. {Bauer}, B.~J. {Bauman}, S.~{Baumont}, E.~{Bechtol}, K.~{Bechtol},
  A.~C. {Becker}, J.~{Becla}, C.~{Beldica}, S.~{Bellavia}, F.~B. {Bianco},
  R.~{Biswas}, G.~{Blanc}, J.~{Blazek}, R.~D. {Blandford}, J.~S. {Bloom},
  J.~{Bogart}, T.~W. {Bond}, M.~T. {Booth}, A.~W. {Borgland}, K.~{Borne}, J.~F.
  {Bosch}, D.~{Boutigny}, C.~A. {Brackett}, A.~{Bradshaw}, W.~N. {Brandt},
  M.~E. {Brown}, J.~S. {Bullock}, P.~{Burchat}, D.~L. {Burke}, G.~{Cagnoli},
  D.~{Calabrese}, S.~{Callahan}, A.~L. {Callen}, J.~L. {Carlin}, E.~L.
  {Carlson}, S.~{Chandrasekharan}, G.~{Charles-Emerson}, S.~{Chesley}, E.~C.
  {Cheu}, H.-F. {Chiang}, J.~{Chiang}, C.~{Chirino}, D.~{Chow}, D.~R. {Ciardi},
  C.~F. {Claver}, J.~{Cohen-Tanugi}, J.~J. {Cockrum}, R.~{Coles}, A.~J.
  {Connolly}, K.~H. {Cook}, A.~{Cooray}, K.~R. {Covey}, C.~{Cribbs}, W.~{Cui},
  R.~{Cutri}, P.~N. {Daly}, S.~F. {Daniel}, F.~{Daruich}, G.~{Daubard},
  G.~{Daues}, W.~{Dawson}, F.~{Delgado}, A.~{Dellapenna}, R.~{de Peyster},
  M.~{de Val-Borro}, S.~W. {Digel}, P.~{Doherty}, R.~{Dubois}, G.~P.
  {Dubois-Felsmann}, J.~{Durech}, F.~{Economou}, T.~{Eifler}, M.~{Eracleous},
  B.~L. {Emmons}, A.~{Fausti Neto}, H.~{Ferguson}, E.~{Figueroa},
  M.~{Fisher-Levine}, W.~{Focke}, M.~D. {Foss}, J.~{Frank}, M.~D. {Freemon},
  E.~{Gangler}, E.~{Gawiser}, J.~C. {Geary}, P.~{Gee}, M.~{Geha}, C.~J.~B.
  {Gessner}, R.~R. {Gibson}, D.~K. {Gilmore}, T.~{Glanzman}, W.~{Glick},
  T.~{Goldina}, D.~A. {Goldstein}, I.~{Goodenow}, M.~L. {Graham}, W.~J.
  {Gressler}, P.~{Gris}, L.~P. {Guy}, A.~{Guyonnet}, G.~{Haller}, R.~{Harris},
  P.~A. {Hascall}, J.~{Haupt}, F.~{Hernandez}, S.~{Herrmann}, E.~{Hileman},
  J.~{Hoblitt}, J.~A. {Hodgson}, C.~{Hogan}, J.~D. {Howard}, D.~{Huang}, M.~E.
  {Huffer}, P.~{Ingraham}, W.~R. {Innes}, S.~H. {Jacoby}, B.~{Jain},
  F.~{Jammes}, M.~J. {Jee}, T.~{Jenness}, G.~{Jernigan}, D.~{Jevremovi{\'c}},
  K.~{Johns}, A.~S. {Johnson}, M.~W.~G. {Johnson}, R.~L. {Jones},
  C.~{Juramy-Gilles}, M.~{Juri{\'c}}, J.~S. {Kalirai}, N.~J. {Kallivayalil},
  B.~{Kalmbach}, J.~P. {Kantor}, P.~{Karst}, M.~M. {Kasliwal}, H.~{Kelly},
  R.~{Kessler}, V.~{Kinnison}, D.~{Kirkby}, L.~{Knox}, I.~V. {Kotov}, V.~L.
  {Krabbendam}, K.~S. {Krughoff}, P.~{Kub{\'a}nek}, J.~{Kuczewski},
  S.~{Kulkarni}, J.~{Ku}, N.~R. {Kurita}, C.~S. {Lage}, R.~{Lambert},
  T.~{Lange}, J.~B. {Langton}, L.~{Le Guillou}, D.~{Levine}, M.~{Liang}, K.-T.
  {Lim}, C.~J. {Lintott}, K.~E. {Long}, M.~{Lopez}, P.~J. {Lotz}, R.~H.
  {Lupton}, N.~B. {Lust}, L.~A. {MacArthur}, A.~{Mahabal}, R.~{Mandelbaum},
  T.~W. {Markiewicz}, D.~S. {Marsh}, P.~J. {Marshall}, S.~{Marshall}, M.~{May},
  R.~{McKercher}, M.~{McQueen}, J.~{Meyers}, M.~{Migliore}, M.~{Miller}, D.~J.
  {Mills}, C.~{Miraval}, J.~{Moeyens}, F.~E. {Moolekamp}, D.~G. {Monet},
  M.~{Moniez}, S.~{Monkewitz}, C.~{Montgomery}, C.~B. {Morrison}, F.~{Mueller},
  G.~P. {Muller}, F.~{Mu{\~n}oz Arancibia}, D.~R. {Neill}, S.~P. {Newbry},
  J.-Y. {Nief}, A.~{Nomerotski}, M.~{Nordby}, P.~{O'Connor}, J.~{Oliver}, S.~S.
  {Olivier}, K.~{Olsen}, W.~{O'Mullane}, S.~{Ortiz}, S.~{Osier}, R.~E. {Owen},
  R.~{Pain}, P.~E. {Palecek}, J.~K. {Parejko}, J.~B. {Parsons}, N.~M. {Pease},
  J.~M. {Peterson}, J.~R. {Peterson}, D.~L. {Petravick}, M.~E. {Libby Petrick},
  C.~E. {Petry}, F.~{Pierfederici}, S.~{Pietrowicz}, R.~{Pike}, P.~A. {Pinto},
  R.~{Plante}, S.~{Plate}, J.~P. {Plutchak}, P.~A. {Price}, M.~{Prouza},
  V.~{Radeka}, J.~{Rajagopal}, A.~P. {Rasmussen}, N.~{Regnault}, K.~A. {Reil},
  D.~J. {Reiss}, M.~A. {Reuter}, S.~T. {Ridgway}, V.~J. {Riot}, S.~{Ritz},
  S.~{Robinson}, W.~{Roby}, A.~{Roodman}, W.~{Rosing}, C.~{Roucelle}, M.~R.
  {Rumore}, S.~{Russo}, A.~{Saha}, B.~{Sassolas}, T.~L. {Schalk},
  P.~{Schellart}, R.~H. {Schindler}, S.~{Schmidt}, D.~P. {Schneider}, M.~D.
  {Schneider}, W.~{Schoening}, G.~{Schumacher}, M.~E. {Schwamb}, J.~{Sebag},
  B.~{Selvy}, G.~H. {Sembroski}, L.~G. {Seppala}, A.~{Serio}, E.~{Serrano},
  R.~A. {Shaw}, I.~{Shipsey}, J.~{Sick}, N.~{Silvestri}, C.~T. {Slater}, J.~A.
  {Smith}, R.~C. {Smith}, S.~{Sobhani}, C.~{Soldahl}, L.~{Storrie-Lombardi},
  E.~{Stover}, M.~A. {Strauss}, R.~A. {Street}, C.~W. {Stubbs}, I.~S.
  {Sullivan}, D.~{Sweeney}, J.~D. {Swinbank}, A.~{Szalay}, P.~{Takacs}, S.~A.
  {Tether}, J.~J. {Thaler}, J.~G. {Thayer}, S.~{Thomas}, A.~J. {Thornton},
  V.~{Thukral}, J.~{Tice}, D.~E. {Trilling}, M.~{Turri}, R.~{Van Berg},
  D.~{Vanden Berk}, K.~{Vetter}, F.~{Virieux}, T.~{Vucina}, W.~{Wahl},
  L.~{Walkowicz}, B.~{Walsh}, C.~W. {Walter}, D.~L. {Wang}, S.-Y. {Wang},
  M.~{Warner}, O.~{Wiecha}, B.~{Willman}, S.~E. {Winters}, D.~{Wittman}, S.~C.
  {Wolff}, W.~M. {Wood-Vasey}, X.~{Wu}, B.~{Xin}, P.~{Yoachim}, and H.~{Zhan}.
\newblock {LSST: From Science Drivers to Reference Design and Anticipated Data
  Products}.
\newblock \emph{The Astrophysical Journal}, 873\penalty0 (2):\penalty0 111,
  Mar. 2019.
\newblock \doi{10.3847/1538-4357/ab042c}.

\bibitem[Izbicki and Lee(2016)]{CDElossIzbicki2016}
R.~Izbicki and A.~B. Lee.
\newblock Nonparametric conditional density estimation in a high-dimensional
  regression setting.
\newblock \emph{Journal of Computational and Graphical Statistics},
  25:\penalty0 1297 -- 1316, 2016.
\newblock URL \url{https://api.semanticscholar.org/CorpusID:2886739}.

\bibitem[Kingma and Ba(2014)]{AdamOptimizerKingma2014}
D.~P. Kingma and J.~Ba.
\newblock Adam: A method for stochastic optimization.
\newblock \emph{CoRR}, abs/1412.6980, 2014.
\newblock URL \url{https://api.semanticscholar.org/CorpusID:6628106}.

\bibitem[Koh and Liang(2017)]{InfluenceKoh2017}
P.~W. Koh and P.~Liang.
\newblock Understanding black-box predictions via influence functions.
\newblock In \emph{International Conference on Machine Learning}, 2017.
\newblock URL \url{https://api.semanticscholar.org/CorpusID:13193974}.

\bibitem[Krishnapriyan et~al.(2021)Krishnapriyan, Gholami, Zhe, Kirby, and
  Mahoney]{CharacteringPINNFailureKrishnapryanMahoney2021}
A.~Krishnapriyan, A.~Gholami, S.~Zhe, R.~Kirby, and M.~W. Mahoney.
\newblock Characterizing possible failure modes in physics-informed neural
  networks.
\newblock In A.~Beygelzimer, Y.~Dauphin, P.~Liang, and J.~W. Vaughan, editors,
  \emph{Advances in Neural Information Processing Systems}, 2021.
\newblock URL \url{https://openreview.net/forum?id=a2Gr9gNFD-J}.

\bibitem[Lang(2014)]{unWiseLang2014}
D.~Lang.
\newblock unwise: Unblurred coadds of the wise imaging.
\newblock \emph{The Astronomical Journal}, 147, 2014.
\newblock URL \url{https://api.semanticscholar.org/CorpusID:119237829}.

\bibitem[Laureijs et~al.(2011)Laureijs, Amiaux, Arduini, Augu{\`e}res,
  Brinchmann, Cole, Cropper, Dabin, Duvet, Ealet, Garilli, Gondoin, Guzzo,
  Hoar, Hoekstra, Holmes, Kitching, Maciaszek, Mellier, Pasian, Percival,
  Rhodes, Criado, Sauvage, Scaramella, Valenziano, Warren, Bender, Castander,
  Cimatti, F{\`e}vre, Kurki-Suonio, Levi, Lilje, Meylan, Nichol, Pedersen,
  Popa, Lopez, Rix, Rottgering, Zeilinger, Grupp, Hudelot, Massey, Meneghetti,
  Miller, Paltani, Paulin-Henriksson, Pires, Saxton, Schrabback, Seidel, Walsh,
  Aghanim, Amendola, Bartlett, Baccigalupi, Beaulieu, Benabed, Cuby, Elbaz,
  Fosalba, Gavazzi, Helmi, Hook, Irwin, Kneib, Kunz, Mannucci, Moscardini, Tao,
  Teyssier, Weller, Zamorani, Osorio, Boulade, Foumond, Giorgio, Guttridge,
  James, Kemp, Martignac, Spencer, Walton, Blumchen, Bonoli, Bortoletto, Cerna,
  Corcione, Fabron, Jahnke, Ligori, Madrid, Martin, Morgante, Pamplona, Prieto,
  Riva, Toledo, Trifoglio, Zerbi, Abdalla, Douspis, Grenet, Borgani, Bouwens,
  Courbin, Delouis, Dubath, Fontana, Frailis, Grazian, Koppenhofer, Mansutti,
  Melchior, Mignoli, Mohr, Neissner, Noddle, Poncet, Scodeggio, Serrano, Shane,
  Starck, Surace, Taylor, Verdoes-Kleijn, Vuerli, Williams, Zacchei, Altieri,
  Sanz, Kohley, Oosterbroek, Astier, Bacon, Bardelli, Baugh, Bellagamba,
  Benoist, Bianchi, Biviano, Branchini, Carbone, Cardone, Clements, Colombi,
  Conselice, Cresci, Deacon, Dunlop, Fedeli, Fontanot, Franzetti, Giocoli,
  Garc{\'i}a-Bellido, Gow, Heavens, Hewett, Heymans, Holland, Huang, Ilbert,
  Joachimi, Jennins, Kerins, Kiessling, Kirk, Kotak, Krause, Lahav, van
  Leeuwen, Lesgourgues, Lombardi, Magliocchetti, Maguire, Majerotto, Maoli,
  Marulli, Maurogordato, McCracken, McLure, Melchiorri, Merson, Moresco,
  Nonino, Norberg, Peacock, Pell{\'o}, Penny, Pettorino, Porto, Pozzetti,
  Quercellini, Radovich, Rassat, Roche, Ronayette, Rossetti, Sartoris,
  Schneider, Semboloni, Serjeant, Simpson, Skordis, Smadja, Smartt, Spano,
  Spiro, Sullivan, Tilquin, Trotta, Verde, Wang, Williger, Zhao, Zoubian, and
  Zucca]{Eulid}
R.~J. Laureijs, J.~Amiaux, S.~Arduini, J.~L. Augu{\`e}res, J.~Brinchmann, R.~E.
  Cole, M.~S. Cropper, C.~Dabin, L.~Duvet, A.~Ealet, B.~Garilli, P.~Gondoin,
  L.~Guzzo, J.~Hoar, H.~Hoekstra, R.~Holmes, T.~D. Kitching, T.~Maciaszek,
  Y.~Mellier, F.~Pasian, W.~J. Percival, J.~D. Rhodes, G.~S. Criado,
  M.~Sauvage, R.~Scaramella, L.~Valenziano, S.~R. Warren, R.~Bender, F.~J.
  Castander, A.~Cimatti, O.~L. F{\`e}vre, H.~Kurki-Suonio, M.~Levi, P.~B.
  Lilje, G.~Meylan, R.~C. Nichol, K.~Pedersen, V.~Popa, R.~R. Lopez, H.-W. Rix,
  H.~J.~A. Rottgering, W.~W. Zeilinger, F.~U. Grupp, P.~Hudelot, R.~J. Massey,
  M.~Meneghetti, L.~Miller, S.~Paltani, S.~Paulin-Henriksson, S.~Pires, C.~J.
  Saxton, T.~Schrabback, G.~Seidel, J.~R. Walsh, N.~Aghanim, L.~Amendola, J.~G.
  Bartlett, C.~Baccigalupi, J.~P. Beaulieu, K.~Benabed, J.-G. Cuby, D.~Elbaz,
  P.~Fosalba, G.~Gavazzi, A.~Helmi, I.~M. Hook, M.~J. Irwin, J.-P. Kneib,
  M.~Kunz, F.~Mannucci, L.~Moscardini, C.~Tao, R.~Teyssier, J.~Weller,
  G.~Zamorani, M.~R.~Z. Osorio, O.~Boulade, J.~J. Foumond, A.~M.~D. Giorgio,
  P.~Guttridge, A.~James, M.~W. Kemp, J.~Martignac, A.~Spencer, D.~J. Walton,
  T.~Blumchen, C.~Bonoli, F.~Bortoletto, C.~Cerna, L.~Corcione, C.~Fabron,
  K.~Jahnke, S.~Ligori, F.~Madrid, L.~Martin, G.~Morgante, T.~Pamplona,
  {\'E}.~Prieto, M.~Riva, R.~Toledo, M.~Trifoglio, F.~Zerbi, F.~B. Abdalla,
  M.~Douspis, C.~Grenet, S.~Borgani, R.~J. Bouwens, F.~Courbin, J.~Delouis,
  P.~Dubath, A.~Fontana, M.~Frailis, A.~Grazian, J.~Koppenhofer, O.~Mansutti,
  M.~Melchior, M.~Mignoli, J.~J. Mohr, C.~Neissner, K.~Noddle, M.~Poncet,
  M.~Scodeggio, S.~Serrano, N.~Shane, J.~L. Starck, C.~Surace, A.~N. Taylor,
  G.~Verdoes-Kleijn, C.~Vuerli, O.~R. Williams, A.~Zacchei, B.~Altieri, I.~E.
  Sanz, R.~Kohley, T.~Oosterbroek, P.~Astier, D.~Bacon, S.~Bardelli, C.~M.
  Baugh, F.~Bellagamba, C.~Benoist, D.~Bianchi, A.~Biviano, E.~Branchini,
  C.~Carbone, V.~F. Cardone, D.~G. Clements, S.~Colombi, C.~J. Conselice,
  G.~Cresci, N.~R. Deacon, J.~S. Dunlop, C.~Fedeli, F.~Fontanot, P.~Franzetti,
  C.~Giocoli, J.~Garc{\'i}a-Bellido, J.~P.~D. Gow, A.~F. Heavens, P.~C. Hewett,
  C.~Heymans, A.~D. Holland, Z.~L. Huang, O.~Ilbert, B.~Joachimi, E.~Jennins,
  E.~Kerins, A.~A. Kiessling, D.~Kirk, R.~Kotak, O.~Krause, O.~Lahav, F.~van
  Leeuwen, J.~Lesgourgues, M.~Lombardi, M.~Magliocchetti, K.~Maguire,
  E.~Majerotto, R.~Maoli, F.~Marulli, S.~Maurogordato, H.~J. McCracken, R.~J.
  McLure, A.~Melchiorri, A.~Merson, M.~Moresco, M.~Nonino, P.~Norberg, J.~A.
  Peacock, R.~Pell{\'o}, M.~T. Penny, V.~Pettorino, C.~D.~N. Porto,
  L.~Pozzetti, C.~Quercellini, M.~Radovich, A.~Rassat, N.~Roche, S.~Ronayette,
  E.~Rossetti, B.~Sartoris, P.~C. Schneider, E.~Semboloni, S.~Serjeant, F.~O.
  Simpson, C.~Skordis, G.~Smadja, S.~J. Smartt, P.~Spano, S.~Spiro,
  M.~Sullivan, A.~Tilquin, R.~Trotta, L.~Verde, Y.~Wang, G.~M. Williger,
  G.~Zhao, J.~Zoubian, and E.~Zucca.
\newblock Euclid definition study report.
\newblock 2011.
\newblock URL \url{https://api.semanticscholar.org/CorpusID:118357133}.

\bibitem[Newman and Gruen(2022)]{ReviewNewman2022}
J.~A. Newman and D.~Gruen.
\newblock Photometric redshifts for next-generation surveys.
\newblock \emph{Annual Review of Astronomy and Astrophysics}, 2022.
\newblock URL \url{https://api.semanticscholar.org/CorpusID:249334966}.

\bibitem[{Oke} and {Gunn}(1983)]{ABMagnitudeSystem}
J.~B. {Oke} and J.~E. {Gunn}.
\newblock {Secondary standard stars for absolute spectrophotometry.}
\newblock \emph{The Astrophysical Journal}, 266:\penalty0 713--717, Mar. 1983.
\newblock \doi{10.1086/160817}.

\bibitem[Papyan et~al.(2020)Papyan, Han, and Donoho]{NeuralCollapsePapayan2020}
V.~Papyan, X.~Han, and D.~L. Donoho.
\newblock Prevalence of neural collapse during the terminal phase of deep
  learning training.
\newblock \emph{Proceedings of the National Academy of Sciences of the United
  States of America}, 117:\penalty0 24652 -- 24663, 2020.
\newblock URL \url{https://api.semanticscholar.org/CorpusID:221172897}.

\bibitem[Park et~al.(2023)Park, Georgiev, Ilyas, Leclerc, and
  Madry]{TRAKPark2023}
S.~M. Park, K.~Georgiev, A.~Ilyas, G.~Leclerc, and A.~Madry.
\newblock Trak: Attributing model behavior at scale.
\newblock In \emph{International Conference on Machine Learning}, 2023.
\newblock URL \url{https://api.semanticscholar.org/CorpusID:257757261}.

\bibitem[{Pasquet} et~al.(2019){Pasquet}, {Bertin}, {Treyer}, {Arnouts}, and
  {Fouchez}]{Pasquet}
J.~{Pasquet}, E.~{Bertin}, M.~{Treyer}, S.~{Arnouts}, and D.~{Fouchez}.
\newblock {Photometric redshifts from SDSS images using a convolutional neural
  network}.
\newblock \emph{Astronomy and Astrophysics}, 621:\penalty0 A26, Jan. 2019.
\newblock \doi{10.1051/0004-6361/201833617}.

\bibitem[Paszke et~al.(2019)Paszke, Gross, Massa, Lerer, Bradbury, Chanan,
  Killeen, Lin, Gimelshein, Antiga, Desmaison, K{\"o}pf, Yang, DeVito, Raison,
  Tejani, Chilamkurthy, Steiner, Fang, Bai, and Chintala]{PyTorchPaszke2019}
A.~Paszke, S.~Gross, F.~Massa, A.~Lerer, J.~Bradbury, G.~Chanan, T.~Killeen,
  Z.~Lin, N.~Gimelshein, L.~Antiga, A.~Desmaison, A.~K{\"o}pf, E.~Yang,
  Z.~DeVito, M.~Raison, A.~Tejani, S.~Chilamkurthy, B.~Steiner, L.~Fang,
  J.~Bai, and S.~Chintala.
\newblock Pytorch: An imperative style, high-performance deep learning library.
\newblock In \emph{Neural Information Processing Systems}, 2019.
\newblock URL \url{https://api.semanticscholar.org/CorpusID:202786778}.

\bibitem[Polsterer et~al.(2016)Polsterer, D'Isanto, and
  Gieseke]{PITAstroPolsterer2016}
K.~L. Polsterer, A.~D'Isanto, and F.~Gieseke.
\newblock Uncertain photometric redshifts.
\newblock \emph{arXiv: Instrumentation and Methods for Astrophysics}, 2016.
\newblock URL \url{https://api.semanticscholar.org/CorpusID:119296950}.

\bibitem[Pruthi et~al.(2020)Pruthi, Liu, Sundararajan, and
  Kale]{TracInGarima2020}
G.~Pruthi, F.~Liu, M.~Sundararajan, and S.~Kale.
\newblock Estimating training data influence by tracking gradient descent.
\newblock \emph{ArXiv}, abs/2002.08484, 2020.
\newblock URL \url{https://api.semanticscholar.org/CorpusID:211204970}.

\bibitem[Raissi et~al.(2019)Raissi, Perdikaris, and Karniadakis]{PINNs}
M.~Raissi, P.~Perdikaris, and G.~Karniadakis.
\newblock Physics-informed neural networks: A deep learning framework for
  solving forward and inverse problems involving nonlinear partial differential
  equations.
\newblock \emph{Journal of Computational Physics}, 378:\penalty0 686--707,
  2019.
\newblock ISSN 0021-9991.
\newblock \doi{https://doi.org/10.1016/j.jcp.2018.10.045}.
\newblock URL
  \url{https://www.sciencedirect.com/science/article/pii/S0021999118307125}.

\bibitem[Sabour et~al.(2017)Sabour, Frosst, and
  Hinton]{CapsuleNetworkSabour2017}
S.~Sabour, N.~Frosst, and G.~E. Hinton.
\newblock Dynamic routing between capsules.
\newblock \emph{ArXiv}, abs/1710.09829, 2017.
\newblock URL \url{https://api.semanticscholar.org/CorpusID:3603485}.

\bibitem[Schmidt et~al.(2020)Schmidt, Malz, Malz, Soo, Almosallam, Brescia,
  Cavuoti, Cohen-Tanugi, Connolly, DeRose, Freeman, Graham, Iyer, Iyer, Jarvis,
  Jarvis, Kalmbach, Kovacs, Lee, Longo, Morrison, Newman, Nourbakhsh, Nuss,
  Pospisil, Tranin, Wechsler, Wechsler, Zhou, Zhou, and
  Izbicki]{PZEvaluationLSSTSchmidt2020}
S.~J. Schmidt, A.~I. Malz, A.~I. Malz, J.~Y.~H. Soo, I.~A. Almosallam,
  M.~Brescia, S.~Cavuoti, J.~Cohen-Tanugi, A.~J. Connolly, J.~DeRose, P.~E.
  Freeman, M.~L. Graham, K.~G. Iyer, K.~G. Iyer, M.~J. Jarvis, M.~J. Jarvis,
  J.~Kalmbach, E.~Kovacs, A.~B. Lee, G.~Longo, C.~B. Morrison, J.~A. Newman,
  E.~Nourbakhsh, E.~Nuss, T.~Pospisil, H.~Tranin, R.~H. Wechsler, R.~H.
  Wechsler, R.~Zhou, R.~Zhou, and R.~Izbicki.
\newblock Evaluation of probabilistic photometric redshift estimation
  approaches for the rubin observatory legacy survey of space and time (lsst).
\newblock \emph{Monthly Notices of the Royal Astronomical Society}, 2020.
\newblock URL \url{https://api.semanticscholar.org/CorpusID:224909363}.

\bibitem[Shah et~al.(2019)Shah, Joshi, Ghosal, Pokuri, Sarkar,
  Ganapathysubramanian, and Hegde]{InvariancesShah2019}
V.~Shah, A.~Joshi, S.~Ghosal, B.~S.~S. Pokuri, S.~Sarkar,
  B.~Ganapathysubramanian, and C.~Hegde.
\newblock Encoding invariances in deep generative models.
\newblock \emph{ArXiv}, abs/1906.01626, 2019.
\newblock URL \url{https://api.semanticscholar.org/CorpusID:174798336}.

\bibitem[Singal et~al.(2021)Singal, Silverman, Jones, Do, Boscoe, and
  Wan]{IdentifyingOutliersSingal}
J.~Singal, G.~Silverman, E.~Jones, T.~Do, B.~M. Boscoe, and Y.~Wan.
\newblock Machine learning classification to identify catastrophic outlier
  photometric redshift estimates.
\newblock \emph{The Astrophysical Journal}, 928, 2021.
\newblock URL \url{https://api.semanticscholar.org/CorpusID:245144697}.

\bibitem[{Strauss} et~al.(2002){Strauss}, {Weinberg}, {Lupton}, {Narayanan},
  {Annis}, {Bernardi}, {Blanton}, {Burles}, {Connolly}, {Dalcanton}, {Doi},
  {Eisenstein}, {Frieman}, {Fukugita}, {Gunn}, {Ivezi{\'c}}, {Kent}, {Kim},
  {Knapp}, {Kron}, {Munn}, {Newberg}, {Nichol}, {Okamura}, {Quinn}, {Richmond},
  {Schlegel}, {Shimasaku}, {SubbaRao}, {Szalay}, {Vanden Berk}, {Vogeley},
  {Yanny}, {Yasuda}, {York}, and {Zehavi}]{SDSS_MGS_2002}
M.~A. {Strauss}, D.~H. {Weinberg}, R.~H. {Lupton}, V.~K. {Narayanan},
  J.~{Annis}, M.~{Bernardi}, M.~{Blanton}, S.~{Burles}, A.~J. {Connolly},
  J.~{Dalcanton}, M.~{Doi}, D.~{Eisenstein}, J.~A. {Frieman}, M.~{Fukugita},
  J.~E. {Gunn}, {\v{Z}}.~{Ivezi{\'c}}, S.~{Kent}, R.~S.~J. {Kim}, G.~R.
  {Knapp}, R.~G. {Kron}, J.~A. {Munn}, H.~J. {Newberg}, R.~C. {Nichol},
  S.~{Okamura}, T.~R. {Quinn}, M.~W. {Richmond}, D.~J. {Schlegel},
  K.~{Shimasaku}, M.~{SubbaRao}, A.~S. {Szalay}, D.~{Vanden Berk}, M.~S.
  {Vogeley}, B.~{Yanny}, N.~{Yasuda}, D.~G. {York}, and I.~{Zehavi}.
\newblock {Spectroscopic Target Selection in the Sloan Digital Sky Survey: The
  Main Galaxy Sample}.
\newblock \emph{The Astronomical Journal}, 124\penalty0 (3):\penalty0
  1810--1824, Sept. 2002.
\newblock \doi{10.1086/342343}.

\bibitem[Stylianou et~al.(2022)Stylianou, Malz, Hatfield, Crenshaw, and
  Gschwend]{LabelNoiseGPZStylianou2022}
N.~Stylianou, A.~I. Malz, P.~W. Hatfield, J.~F. Crenshaw, and J.~Gschwend.
\newblock The sensitivity of gpz estimates of photo-z posterior pdfs to
  realistically complex training set imperfections.
\newblock \emph{Publications of the Astronomical Society of the Pacific}, 134,
  2022.
\newblock URL \url{https://api.semanticscholar.org/CorpusID:247154897}.

\bibitem[Szegedy et~al.(2015)Szegedy, Vanhoucke, Ioffe, Shlens, and
  Wojna]{InceptionSzegedy2015}
C.~Szegedy, V.~Vanhoucke, S.~Ioffe, J.~Shlens, and Z.~Wojna.
\newblock Rethinking the inception architecture for computer vision.
\newblock \emph{2016 IEEE Conference on Computer Vision and Pattern Recognition
  (CVPR)}, pages 2818--2826, 2015.
\newblock URL \url{https://api.semanticscholar.org/CorpusID:206593880}.

\end{thebibliography}
